\documentstyle[aaspp4]{article}
\begin{document}

\title{AN ORIGIN OF THE HUGE FAR-INFRARED LUMINOSITY OF STARBURST MERGERS}

\author{Yoshiaki Taniguchi and Youichi Ohyama}

\affil{Astronomical Institute, Tohoku University, Aoba, Sendai 980-8578, Japan}


\begin{abstract}
Recently Taniguchi and Ohyama found that the higher $^{12}$CO to $^{13}$CO
integrated intensity ratios at a transition $J$=1--0,
$R = I(^{12}$CO)$/I(^{13}$CO) $\gtrsim 20$, in  a sample of 
starburst merging galaxies such as Arp 220 are mainly attributed to 
the depression of $^{13}$CO emission with respect to $^{12}$CO.
Investigating the same sample of galaxies analyzed by Taniguchi \& Ohyama,
we find that there is a tight, almost linear correlation between the dust mass and
$^{13}$CO luminosity.
This implies that dust grains are also depressed in the high-$R$ starburst 
mergers, leading to the higher dust temperature ($T_{\rm d}$) in them
because of the relative increase in the radiation density.
Nevertheless, the average dust mass ($M_{\rm d}$) of the high-$R$ starburst 
mergers is higher significantly than that of non-high $R$ galaxies.
This is naturally understood because the galaxy mergers could accumulate a lot of
dust grains from their progenitor galaxies together with supply of 
dust grains formed newly in the star forming regions.
Since $L$(FIR) $\propto M_{\rm d} T_{\rm d}^5$
given the dust emissivity law, $S_\nu \propto \lambda^{-1}$, the increases in both
$M_{\rm d}$ and $T_{\rm d}$ explain well why the starburst 
mergers are so bright in the FIR. We discuss that the superwind activity plays
an important role in destroying dust grains as well as dense gas clouds
in the central region of mergers.
\end{abstract}




\section{INTRODUCTION}

It is known that some starburst galaxies have significantly high
$R = ^{12}$CO($J$=1--0)$/^{13}$CO($J$=1--0) integrated line intensity 
ratios, $R \gtrsim 20$,  while typical galaxies have $R \simeq 10$ 
(Aalto et al. 1991, 1995; Casoli, Dupraz, \& Combes 1992a, 1992b;
Turner \& Hurt 1992; Garay, Mardones, \& Mirabel 1993; Henkel et al. 1998).
Recently Taniguchi \& Ohyama (1998; hereafter TO98) have  shown 
that the observed higher $R$ values are mainly attributed to
the lower intensity in $^{13}$CO($J$=1--0) with respect to $^{12}$CO($J$=1--0).
This finding suggests that the $^{13}$CO depression provides an  
important clue to the understanding physical properties of the interstellar
medium in the starburst galaxies because these properties should be
controlled by the star formation histories in them.
Interestingly, almost all the high-$R$ starburst galaxies show morphological 
evidence for major mergers and their far-infrared luminosities are generally high;
e.g., $L$(FIR) $\gtrsim 10^{11} L_\odot$ (TO98). 
Among the high-$R$ starburst galaxies, Arp 220 indeed belongs to a class of 
ultraluminous infrared galaxies (hereafter ULIGs, 
Sanders et al. 1988; see for a review Sanders \& Mirabel 1996).
Therefore, the origin of the high-$R$ starburst galaxies seems to be 
physically related to that of ULIGs. 
In this Letter, in order to understand the origin of huge FIR luminosities 
of the high-$R$ starburst galaxies, 
we present a statistical analysis for the galaxy sample studied by TO98.


\section{PHYSICAL PROPERTIES OF THE INTERSTELLAR MEDIUM IN THE STARBURST MERGERS}

All the data analyzed here are taken from TO98.
The TO98 sample contains eight high-$R$ starburst mergers with $R \gtrsim 20$ 
(hereafter high-$R$ mergers) and 43 non-high-$R$ galaxies  with $R \simeq 10$
(hereafter normal-$R$ galaxies). 
The high-$R$ starburst mergers are NGC 1614, NGC 3256, NGC 4194, NGC 6240,
Arp 220, Arp 236 (ESO 541--IG 23), Arp 299, and IRAS 18293$-$3413.
Among the 43 normal-$R$ galaxies, two galaxies, NGC 55 and NGC 404, are 
dwarf galaxies. Therefore, we omit them in the later 
statistical analysis. The galaxies analyzed here were selected solely by 
a criterion that both $^{12}$CO($J$=1--0) and $^{13}$CO($J$=1--0) data are 
available in the literature. Therefore our sample is not a statistically complete one 
in any sense.
However, this is enough to investigate the difference between the high-$R$
mergers and the normal-$R$ galaxies. 
Note also that some starburst galaxies such as NGC 253
are included in the normal-$R$ sample. This suggests that the observed
high-$R$ values in the starburst mergers are not due to the starburst activity 
itself and thus some other mechanisms should be responsible for them.

For the remaining 49 galaxies, we derive dust temperatures ($T_{\rm d}$)
using IRAS 60$\mu$m and 100$\mu$m fluxes (Moshir et al. 1992).
Here we assume  the dust emissivity law, $S_\nu \propto \lambda^{-1}$.
We also estimate the mass of dust which is responsible for the radiation
at  60$\mu$m and 100$\mu$m from
$M_{\rm d} = (4 a \rho/3) D^2 S_{100} [Q_{100} B_{100}(T)]^{-1}$
where $a$ is the grain radius, $\rho$ is the grain density,
$D$ is the distance of galaxy, $S_{100}$ is the IRAS 100$\mu$m flux, 
$Q_{100}$ is the grain emissivity at 100 $\mu$m, and $B_{100}(T)$ is the 
value of Planck function at 100 $\mu$m (e.g., Hildebrand 1983).
Following Devereux \& Young (1990), we use a simplified formula
$M_{\rm d} = C {S_{100}} D^2 (e^{144/T_{\rm d}} - 1) (M_\odot)$ 
where $C$ is the grain opacity. Adopting 
$3 Q_{100}/(4 a \rho) = 25$ cm$^2$ g$^{-1}$ (Hildebrand 1983),
we use $C = 4.58$ in our analysis. The distance is in units of Mpc.
Statistical properties of the sample are summarized in Table 1.
The high-$R$ mergers have both brighter
$L$(FIR) and $L(^{12}$CO) for their $M_{\rm d}$.
Indeed the average $L$(FIR) and $L(^{12}$CO) of the high-$R$  mergers
are higher by  factors of 20.0 and 10.7, respectively,
than those of the normal-$R$ galaxies. On the other hand, 
the average $L(^{13}$CO) and $M_{\rm d}$ of the high-$R$ mergers
are higher than those of the normal-$R$ galaxies only by factors of 4.4 and 6.3,
respectively. 

In Figure 1, we show diagrams of $L(^{12}$CO), $L(^{13}$CO), and $M_{\rm d}$
against $L$(FIR). As shown by TO98, although there is a tight correlation
between $L(^{12}$CO) and $L$(FIR), $L(^{13}$CO) becomes lower with
increasing $L$(FIR). The bottom panel of Figure 1 shows that the dust mass also
becomes lower with increasing $L$(FIR), suggesting that the dust content in the
high-$R$  mergers is more intimately related to the $^{13}$CO gas content.
In order to confirm this, 
we show diagrams of $L(^{12}$CO) and $L(^{13}$CO) against $M_{\rm d}$ in Figure 2.
We find that  there is a tight, almost linear  correlation
between $M_{\rm d}$ and $L(^{13}$CO) for all the galaxies analyzed here;
log $L$($^{13}$CO) = (0.942$\pm$0.064) log $M_{\rm d} + (1.138\pm0.288)$
with a correlation coefficient of 0.92. 
Therefore, we conclude that dust grains are more intimately associated
with $^{13}$CO gas rather than with $^{12}$CO gas and that both 
the $^{13}$CO gas and dust grains are depressed in the high-$R$ mergers. 
Although the average dust mass of the high-$R$ mergers is higher 
by a factor of 6.3 than that of the normal-$R$ galaxies,  
this average dust mass is not as high as that expected from the higher 
$^{12}$CO luminosities of the high-$R$ mergers;
the average dust mass of the high-$R$ mergers is lower
by a factor of 1.7 than that expected if their
$M_{\rm d}/L(^{12}$CO) ratios are nearly the same as those of
the normal-$R$  galaxies. 

\section{DISCUSSION}

\subsection{Effect of the Dust Depression on $L$(FIR)}

The present analysis has shown that both the dust and the $^{13}$CO gas are
more depressed in the high-$R$ mergers with respect to the $^{12}$CO gas.
We consider a possible effect of the dust depression 
on the $L$(FIR) in the high-$R$ mergers.
Since the FIR emission is generally believed to arise from large grains
which are in thermal equilibrium with the ambient radiation field
(e.g., Mouri, Kawara, \& Taniguchi 1997 and references therein),
the dust temperature is related to the radiation density ($U_{\rm rad}$)
such that $T_{\rm d} \propto U_{\rm rad}^{1/4}$ (e.g., Spitzer 1978).
Since the average dust temperatures are 43.7 K and 35.4 K
for the high-$R$ mergers and the normal-$R$ galaxies, respectively
(Table 1), the radiation density in the high-$R$ mergers is higher
by a factor of 2.3 than that in the normal-$R$ galaxies.
Since $U_{\rm rad} \sim L_{\rm rad}/M_{\rm d}$ where $L_{\rm rad}$ is the
luminosity responsible for heating dust grains,
the relative depression in $M_{\rm d}$ leads to 
the increase in $U_{\rm rad}$ by a factor of 1.7, which can almost explain
the relative higher radiation density in the high-$R$ mergers.
Therefore, it is suggested that the higher radiation density in the
starburst mergers is achieved by the relative depression of dust grains.

The FIR luminosity is proportional to the product of the dust mass
and a high power of the dust temperature such that $L$(FIR)
$\propto M_{\rm d} T_{\rm d}^5$ given the dust emissivity law,
$S_\nu \propto \lambda^{-1}$.
The average dust mass of the high-$R$ mergers is higher by a factor of 6.3
than that of the normal-$R$ galaxies.
This higher dust mass can be attributed to the accumulation from
the progenitor galaxies because the starburst mergers should come from
mergers between or among gas-rich galaxies
(Sanders et al. 1988; Taniguchi \& Shioya 1998).
Also, it is expected that dust grains could form efficiently
in the intense star forming regions (e.g., Wang 1991).
However, the huge FIR luminosities of the high-$R$ mergers cannot be 
explained solely by the higher dust mass in them because the average 
$L$(FIR) of the high-$R$ mergers is higher by a factor of 20 than that 
of the normal-$R$ galaxies (Table 1).
Therefore we have to take account of another factor, $T_{\rm d}^5$, which 
enhances the average FIR luminosity of the high-$R$ mergers by a factor of 
$(43.7/35.4)^5 \simeq 2.9$  with respect to that of the normal-$R$ galaxies
although the increase in $M_{\rm d}$ is primarily important.
Accordingly, the increases in both $ M_{\rm d}$ and  $T_{\rm d}$ are indeed responsible
for the enhancement of the FIR luminosity of the high-$R$ mergers on the average
by a factor of 18 ($\simeq  6.3 \times 2.9$) with respect to the normal-$R$ galaxies,
being comparable to the actual difference ($\simeq 20$) between them. 

\subsection{What Destroy Dust Grains in the High-$R$ Mergers ?}

We consider why dust grains are depressed in the high-$R$ mergers.
Galaxy mergers cause efficient gas fueling toward the nuclear regions 
of the merging systems that ultimately can trigger intense starbursts,
either as a result of the piling of gas (e.g., Barnes \& Hernquist 1991;
Mihos \& Hernquist 1994) or by the dynamical effect of supermassive binary
black holes  (Taniguchi \& Wada 1996).
Multiple mergers are also responsible for
the formation of starburst mergers (Taniguchi \& Shioya 1998). 
The dense gas clouds in which the majority of dust grains may be present
were consumed in the past starburst activity.
This may explain partly the relative depression of both dense gas traced
by $^{13}$CO and dust grains. Another key process is the so-called
superwind which is a  blast wave driven by a collective effect of a 
large number of supernovae (Heckman, Armus, \& Miley 1987, 1990; 
Heckman et al. 1996; Taniguchi, Trentham, \& Shioya 1998).
Since it is generally considered that dust grains in the interstellar medium
are destroyed mainly by fast shocks driven by supernova explosions 
through  processes of both sputtering and grain-grain collisions 
(e.g., McKee 1989), it is very likely that superwinds occurring
shortly after the starbursts lead to the destruction of dust grains.

We investigate whether or not there is evidence for shock heating
driven by superwinds in the high-$R$ mergers studied here.
In order to examine the predominance of shock heating,
we show an optical excitation diagram of [N {\sc ii}]$\lambda$6583/H$\alpha$ vs. 
[O {\sc iii}]$\lambda$5007/H$\beta$ (Veilleux \& Osterbrock 1987) in Figure 3.
The optical emission-line data are taken from Armus et al. (1989) and 
Kim et al. (1995). The three high-$R$ mergers (NGC 1614, NGC 6240, and  Arp 220)
show LINER-like excitation. NGC 4194 and Arp 299 are also close to the domain
of the LINER. The recent mid-infrared spectroscopy
by Genzel et al. (1998) suggests that NGC 6240 and Arp 220 are star-formation
dominated objects. Therefore, it is unlikely that their optical emission-line
nebulae are powered by the putative central engine of active galactic
nuclei in them.
Thus their LINER-like excitation conditions can be attributed to 
shock heating (see also Veilleux et al. 1995; Kim et al. 1998; 
Taniguchi et al. 1998). In order to examine if the LINER-like excitation can
be explained by shock heating, we show results of shock heating models by
Dopita \& Sutherland (1995)  in Figure 3. 
The excitation properties of both Arp 220 and NGC 6240 can be explained 
by fast shock models with the shock velocity of $\sim$ 300 -- 400 km s$^{-1}$.
Since the H$\alpha$ line widths of these galaxies are $\sim$ 600 km s$^{-1}$
and $\sim$ 900 km s$^{-1}$, respectively (Armus et al. 1989), the shock
interpretation seems reasonable. We also note that the remaining H {\sc ii}
region like objects have narrower line widths, e.g., a few 100 km s$^{-1}$
(Armus et al. 1989). Therefore, these objects may be in late phases of the 
superwind activity; e.g., a few 10$^8$ years after the onset of the 
superwind. Since the time scale of injection of dust grains from stars
is $\sim 10^9$ years (McKee 1989), these galaxies may still have
relatively small amount of dust grains. 

Heckman et al. (1987) first noted that starburst galaxies with evidence for
superwinds have unusual IRAS colors; $\alpha(25,60) < -1.5$ and 
$\alpha(60,100) > -0.5$ where $\alpha(\lambda_1, \lambda_2) = -$
log[$S_\nu(\lambda_1)/S_\nu(\lambda_2)$]/log($\lambda_1/\lambda_2)$
(see also Heckman et al. 1990).
The higher dust temperatures of the high-$R$ mergers explain
the condition $\alpha(60,100) > -0.5$, corresponding to $T_{\rm d} \gtrsim$ 42 K 
under the dust emissivity law, $S_\nu \propto \lambda^{-1}$. 
On the other hand, if small grains are destroyed efficiently by the superwinds,
together with possible extinction even at 25 $\mu$m (e.g., Condon et al. 1991),
IRAS 25 $\mu$m emission should be quenched significantly,
resulting in the condition of $\alpha(25,60) < -1.5$.
Although we cannot give any quantitative discussion here,
the above consideration explains the unusual IRAS colors
of the superwind galaxies qualitatively.

Finally we consider why the observational properties of 
the high-$R$ mergers are different significantly from those
of other star-forming galaxies. One interesting property of
the starbursts in galaxy mergers is that the ages of the starbursts
are often estimated as old as $\sim 10^8$ years, being older than 
the typical starburst,
$\sim 10^7$ years (e.g., Mouri \& Taniguchi 1992; Prestwich, Joseph,
\& Wright 1994). Further, poststarburst regions traced by 
a large number of A-type stars are often found
in the circumnuclear regions of the mergers (Armus, Heckman, \& Miley 
1989; Larkin et al. 1995). These suggest that the superwind activity 
could occur in some  recent epochs in the high-$R$ mergers;
e.g., $\sim 3 \times 10^7$ years ago for Arp 220 (Heckman et al. 1996).
Another important property of the high-$R$ mergers is that
a huge amount of molecular gas is concentrated in their  central region
(Scoville et al. 1991, 1997; Downes \& Solomon 1998).
This means that starbursts occurred in the very gas-rich environment. In fact,
the molecular gas surface density in the central region 
of Arp 220, $\sigma_{\rm H_2} \sim 10^4$ cm$^{-2}$,
is higher by a few orders of magnitudes than those in typical starburst galaxies
such as IC 342 (Kennicutt 1998).  Thus the dense gas clouds in the high-$R$
mergers are inevitably affected by the superwind activity
(Taniguchi, Trentham, \& Shioya 1998),
leading to the depression of both dense gas and dust grains.
In summary, the long-lasting starbursts in the very gas-rich
environment make the high-$R$ mergers a particular class of 
starburst galaxies with superwinds.

\vspace{1ex}

We would like to thank an anonymous referee for useful comments and
suggestions. 
Y.O. was supported by the Grant-in-Aid for JSPS Fellows by
the Ministry of Education, Science, Sports and Culture.
This work was supported in part by the Ministry of Education, Science,
Sports and Culture in Japan under Grant Nos. 07055044, 10044052, and 10304013.


\newpage
\begin{deluxetable}{cccc}
\tablewidth{39pc}
\tablecaption{Comparisons of physical properties between the high-$R$ mergers
          and normal-$R$ galaxies}
\tablehead{
\colhead{} &
\colhead{Units} &
\colhead{Starburst mergers} &
\colhead{Normal galaxies}
}
\startdata
$R$   &     & 27.9$\pm$10.6  & 11.3$\pm$3.3 \nl
log $L(^{12}$CO) & ($L_\odot$) & 9.42$\pm$0.33 & 8.39$\pm$0.63 \nl
log $L(^{13}$CO) & ($L_\odot$) & 8.00$\pm$0.41 & 7.36$\pm$0.60 \nl
log $L$(FIR) & ($L_\odot$) & 11.39$\pm$0.28 & 10.09$\pm$0.59 \nl
log $M_{\rm d}$ & ($M_\odot$) & 7.38$\pm$0.28 & 6.58$\pm$0.52 \nl
$T_{\rm d}$ & (K) & 43.7$\pm$2.6 & 35.4$\pm$4.2  \nl
\enddata
\end{deluxetable}

\newpage

\begin{figure}
\epsfysize=18.5cm \epsfbox{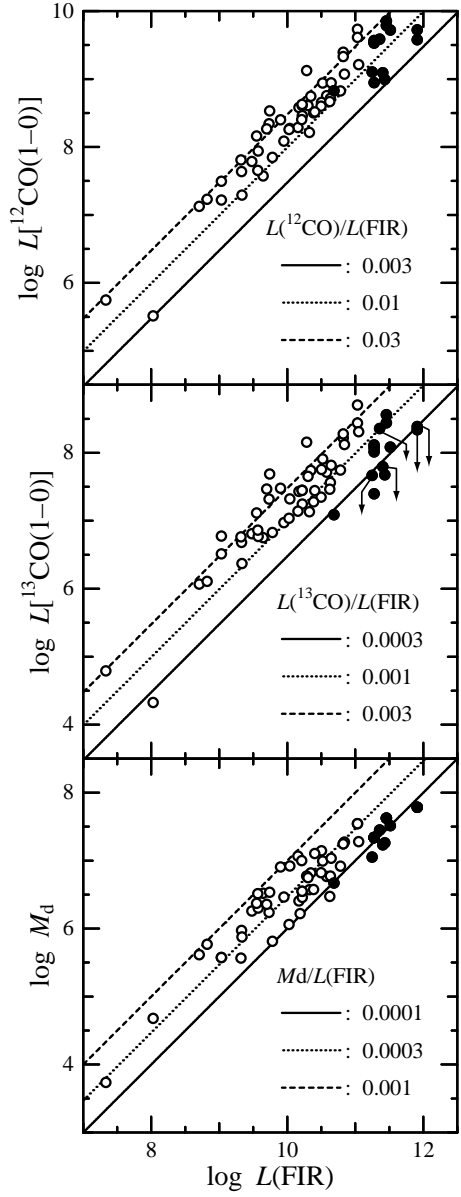}
\caption[]{
Diagrams of $L$[$^{12}$CO($J$=2--1)] (top),
and $L$[$^{13}$CO($J$=2--1)] (middle), and $M_{\rm d}$ (bottom) 
against $L$(FIR). 
The high-$R$ mergers are shown by filled circles while the remaining
normal-$R$  galaxies are shown by open ones.
\label{fig1}
}
\end{figure}

\begin{figure}
\epsfysize=18.5cm \epsfbox{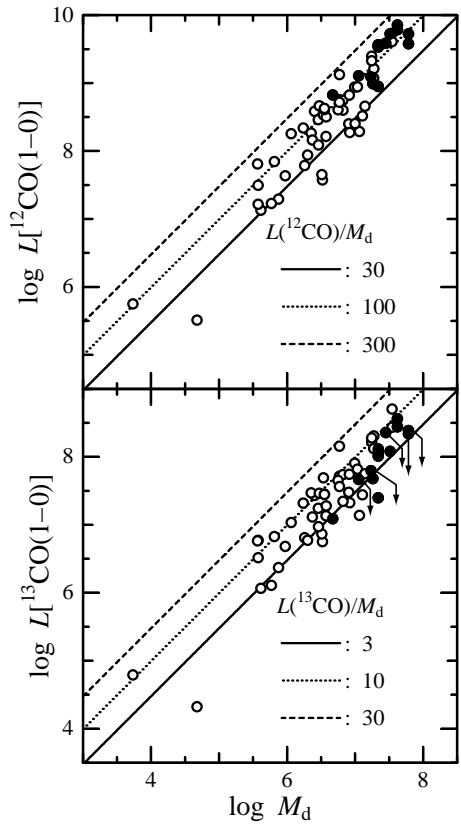}
\caption[]{
Diagrams of $L$[$^{12}$CO($J$=2--1)] (upper)
and $L$[$^{13}$CO($J$=2--1)] (lower) against $M_{\rm d}$.
The high-$R$ mergers are shown by filled circles while the remaining
normal-$R$ galaxies are shown by open ones.
\label{fig2}
}
\end{figure}

\begin{figure}
\epsfysize=15cm \epsfbox{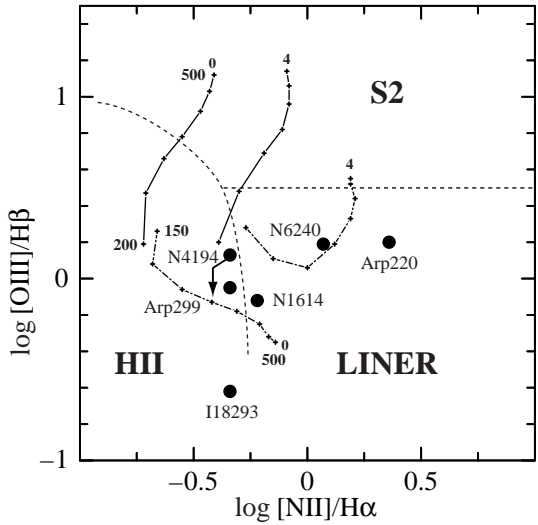}
\caption[]{
Excitation diagram of [N {\sc ii}]$\lambda$6583/H$\alpha$ vs. 
[O {\sc iii}]$\lambda$5007/H$\beta$ for the high-$R$ mergers.
Dashed curve shows the distinction among H {\sc ii} region-like objects, 
Type 2 Seyferts (S2), and LINERs, taken from Veilleux et al. (1995).
The shock models overlaid in the diagrams are taken from 
Dopita \& Sutherland (1995); 1) ^^ ^^ shock + precursor" models (solid
curves) with magnetic parameter $B/n^{1/2}$ = 0 and 4 $\mu$G cm$^{3/2}$
and shock velocity from 200 km s$^{-1}$ to 500 km s$^{-1}$, and 
2) ^^ ^^ shock only" models (dash-dot
curves) with magnetic parameter $B/n^{1/2}$ = 0 and 4 $\mu$G cm$^{3/2}$
and shock velocity from 150 km s$^{-1}$ to 500 km s$^{-1}$.
\label{fig3}
}
\end{figure}

\end{document}